# Evolution of Accreting Binary Systems on the Spin-up Line

[1]Ali Taani, [2]Mashhoor A. Al-Wardat and [3]Awni Khasawneh

[1] *Applied Science Department, Aqaba University College, Al-Balqa' Applied University, P.O. Box 1199, 77110 Aqaba, Jordan*
[2]*Department of Physics, Al-Hussein Bin Talal University, P.O.Box 20, 71111, Ma'an, Jordan*
[3]*The Royal Jordanian Geographic Center, 11941, Amman, Jordan*

**A B S T R A C T**
The measured characteristics of binary pulsars provide valuable insights into the evolution of these systems. We study the aspects of binary evolution particularly relevant to binary Millisecond Pulsars (MSPs), and the formation of close binaries involving degenerate stars through a spin-evolution diagram (spin-up line). For this task, we use a wide variety of binaries, including those with compact components that observed in different energy bands, which we analyze them according to the spin-up line. Their formation and evolution over timescales of binary evolution models are investigated in order to grab any constraint on their evolution, and to estimate the masses of neutron stars with different mass-transfer histories.

## INTRODUCTION

The last decade has seen a dramatic growth in our understanding of compact object populations, almost all driven by new facilities such as the Hubble Space Telescope and the ATNF pulsar catalogue for radio pulsar work, in addition to Chandra and XMM-Newton observatories for studying accreting neutron stars (NSs) and black holes. Pulsars are classified based on their observed properties. Their spin period (*P*) and magnetic field strength (*B*) together provide formation and evolution mechanisms of binary & Millisecond Pulsars (MSPs). Subsequent to the pulsar discoveries, we searched for archival X-ray, Gamma-ray and radio observations covering the new pulsars' sky locations. For the current analysis of Gamma-ray pulsars, we use LAT data (The Second Fermi Large Area Telescope Catalog of Gamma-ray Pulsars). The population is subject to very different selection effects than the general radio pulsar population (Saz Parkinson *et al*. 2010). Fig. 1 shows that young pulsars have spin periods of about 1 second and magnetic fields strengths of about $10^{12}$ G (Lorimer 2008).

Pulsars are usually isolated objects, while binary MSPs represent the end point of stellar evolution, and their observed orbital and stellar properties are fossil records of their evolutionary history. Thus, one can use binary pulsar systems as key probes of stellar astrophysics (Stairs 2004). From the sky distribution in Galactic coordinates detected in various surveys of pulsars shown in Fig. 2. Observations of such objects occur primarily at radio wavelengths. The radio and Fermi pulsars are distributed closely on the galactic plane. Therefore, factors which limit the region of the Galaxy observed (or considered), such as the numerous selection effects which occur in radio pulsar observations, can modify the underlying pulsar population scale heights within -z < 10 kpc of the Galactic plane (as discussed by many works including Taylor & Manchester (1977) and Narayan & Ostriker (1990). While in the distribution of MSPs are likely to be located where they formed, hence the majority of MSPs have been found in the core of the galaxy and tend to diffuse into the spiral arms. The difference simply reflects the observational bias against detecting short-period pulsars with increasing distance from the Sun. This is one of many selection effects that pervade the observed sample (Lorimer 2008). Since all binary pulsars which had a history of mass accretion (so-called 'recycled') tend to have much weaker magnetic fields than normal isolated pulsars (Alpar *et al*. 1982, Taani et al. 2012a,b, Taani et al. 21013), it is thought that accretion in some way causes a weakening of the surface dipole magnetic field of NSs (Taam & van den Heuvel 1986) and several theories have been put forward to explain this accretion-induced field decay Bhattacharya & van den Heuvel 1991; van den Heuvel 2004; Wang *et al*. 2011). NSs accreting materials from their low-mass binary companions are spun up by the angular momentum carried by the accreted materials while the magnetic field decays. The X-ray binaries are the evolutionary progenitors to these 'recycled' MSPs. It is evident that the *B* and *P* of X-ray pulsars and recycled pulsars are correlated with the duration of both the accretion phase and the total amount of matter accreted (Wang *et al*. 2011). We are mainly focusing on the formation and evolution of MSPs and binary pulsars starting from the point where the NS has already formed, and then during the phases of accretion till the spin-up. We provide an updated distribution to the wide variety and natures of binaries with a compact component. In addition, we performed a statistical study of these binary populations including normal pulsars, MSPs, Fermi pulsars, Supernova Remnants (SNRs) and magnetars that are observed in different energy bands, which we plot together in the magnetic field versus spin period diagram.

*Formation and Evolution of Accreting Neutron Stars:*

**Corresponding Author:** Ali Taani, Applied Science Department, Aqaba University College , Al-Balqa' Applied University, P.O. Box 1199, 77110 Aqaba, Jordan.
E-mail: ali.taani@bau.edu.jo, ali82taani@gmail.com

The evolution of the surface dipole magnetic field strength *B* versus spin period *P* (so-called spin-up line or *B-P* diagram), for a NS that is born in a close binary system as measured from all available radio, X-ray, and Gamma observations is depicted in Fig. 1. Pulsars are born in the upper left part of the diagram with high magnetic field strength ($B \sim 10^{12}$ G) and extreme rapid spin ~ 30 ms. As such, they will within a few million years spin-down and move along a horizontal track eventually crossing the 'death-line'. No radio pulsars can exist to the right of this line as the polar cap electric field has become too weak to create pairs (Bhattacharya & van den Heuvel 1991), and no more pulsar wind can be produced causing the radio pulsar emission to cease. The region to the right of the death-line is the graveyard. During their stay in the graveyard the system further spins down as the weak winds of the companion star (or other processes) may make the magnetic dipole moment decay via mass transfer, causing it to move downwards in the *B-P* diagram (see schematic in Fig. 3). After some time, the accretion of matter provides angular momentum to the NSs causing them to spin-up with shorter spin periods meaning they move towards the left in the *B-P* diagram until they reach the spin-up line. Later after the companions have themselves become compact stars (white dwarfs or NSs) or have disappeared, the spun-up NSs become observable as MSPs. Eventually they will slowly spin down, that is, once again move to the right in the diagram. Examples of evolutionary paths are expected to be found only to the right of the spin-up line and to the left of the death line. As we can note in the *B-P* diagram, the connection between high-B pulsars and magnetars (top-right corner) is strengthened by the discovery of a low-*B* SGR (PSR B1509-58) which shows magnetar-like activity despite its canonical magnetic field strength (Rea *et al*. 2010). While pulsars emit steady, beamed electromagnetic radiation from radio up to high energies, ultimately via the loss of rotational kinetic energy ($\dot{E} = 4\pi^2 \dot{P}/P^3$ erg s$^{-1}$), magnetars dissipate their extremely high surface magnetic fields ($B \sim 10^{14} - 10^{15}$ G) in luminous X-ray emission. Hence, about 20 % of all NSs are formed as Magnetars (Perna & Pons 2011). On the other hand, young pulsars associated with SNRs appear to be born with reasonable small periods, $P \leq 0.1$ s, and strong magnetic field $B \sim 10^{12}$ G, which can provide independent distance and age estimates for both objects and, with a statistically significant sample of associations, constraints on the birth properties of NSs, including initial period, and magnetic field, can be found. Although the comparatively short lifetimes of SNRs mean that the number of pulsar-SNR associations is quite small. However the rare associations are of high interest, as they constrain a number of pulsar and SNR parameters. PSR J1846-0258 was discovered in the Rossi X-ray Timing Explorer data RXTE with $B_s = 4.9 \times 10^{13}$ G (Gotthelf *et al*. 2000). Its spin period of 324 ms and its spin-down rate of $\dot{P} = 7.1 \times 10-12$ ss$^{-1}$ indicate this pulsar is one of the youngest known ($t_c = P/2\dot{P}$) ($t_c = 723$ yr). However, PSRJ1119 - 6127 and PSR B1509-58 suggest the existence of a class of equally young pulsars but with higher magnetic fields, longer periods, and much lower spin-down luminosities (Pivovaroff *et al*. 2001). Our sample of Fermi pulsars are divided into millisecond and normal gamma-ray pulsars. Consequently the observed MSP gamma-ray profiles and their relation to the radio profiles are similar to those observed for young pulsars (Champion *et al*. 2008).

*Conclusions:*

The knowledge of the field of the vast population of active normal, MSPs and their possible progenitors, as well as of the physical processes involved in the formation and evolution of NSs and their magnetic fields, allows us to probe the underlying NSs in different natures of binaries with a compact component evolution distribution in detail. Subsequently poses challenges in attaining deep analysis of their evolution track. The updated parameters for the binary systems studied here, together with recently discovered similar systems, allow us to update previous distributions along the spin-up line. This diagram represents the minimum spin period to which such a spin up may proceed in an Eddington-limited accretion scenario. Therefore, the observed properties of binary pulsars reveal much information about the evolutionary history of different types of binaries (including their spin periods, magnetic field, component masses, and supernova processes) due to the conversion of angular momentum and mass transfer. We hope that observations in most pulsars with both gamma-ray and radio observations by more sensitive instruments on future missions, like ASTROSAT (Paul & The LAXPC Team 2009) and FAST (Wang 2006), will provide important clues regarding the pulsars formation mechanism and distributions.

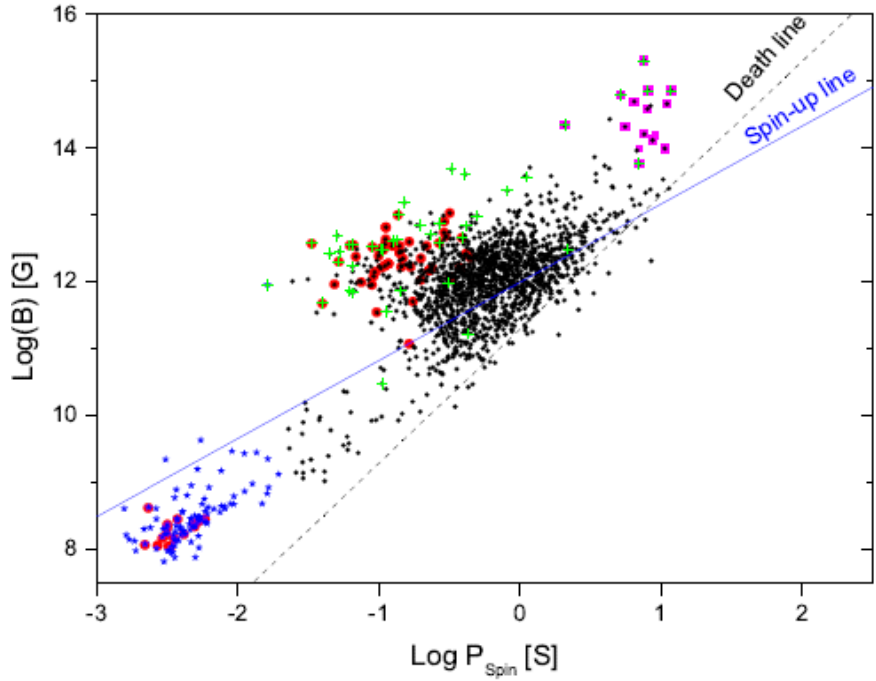

**Fig. 1:** This is a scatter plot of magnetic field and spin period for pulsars in the current (March 2012) public ATNF catalogue (Manchester *et al.* 2005). MSPs (together with binaries) are marked with blue stars, Fermi pulsars (red circle), SNRs (pink stars) and magnetars (green triangle). The solid line indicates the spin-up line which denotes the terminal period for objects moving to the left in the diagram owing to accretion that become the recycled pulsars once accretion ceases. The death line (shaded line) marks the locations where radio pulsars emit less radiation or turn off entirely (Bhattacharya & van den Heuvel 1991).

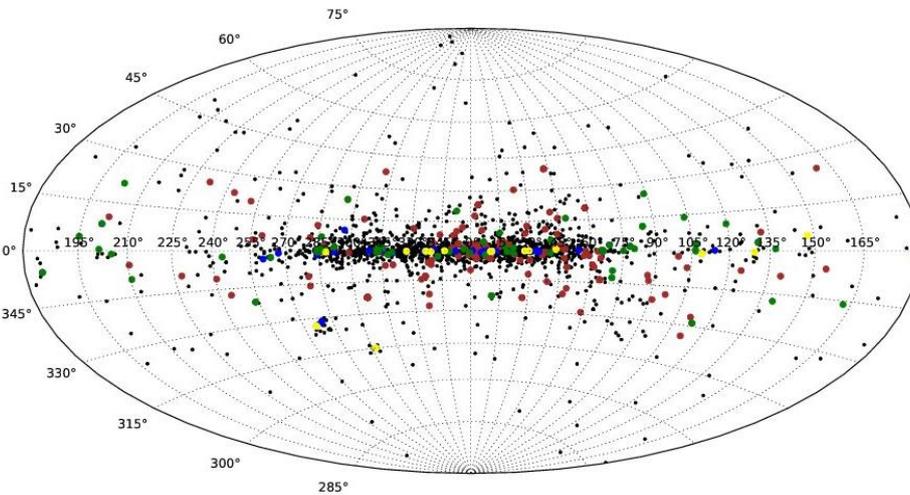

**Fig. 2:** The full-sky distribution of all known pulsars in different nature, shown in Galactic coordinates in which the plane of the Galaxy is at the equator, and the Galactic center is at the origin. (Blue is the ATNF binary pulsars, black is the Fermi pulsars, red is the HMXBs, green is the LMXBs, Stars are isolated ATNF MSPs, triangle is the ATNF MSPs).

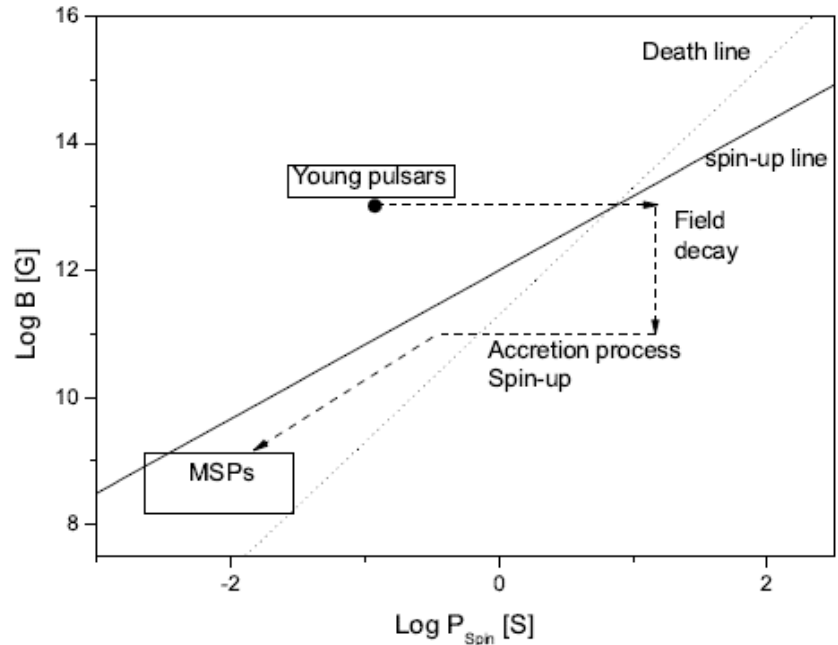

**Fig. 3:** Evolutionary track of MSPs on the *B-P* diagram.

## ACKNOWLEDGMENTS

We would like to thank Prof. Chengmin Zhang for the discussion and suggestions.
This research was written in part during my research at National Astronomical Observatories (NAOC).